\documentclass[apj]{emulateapj}

\slugcomment{accepted for publication in ApJ}
\shorttitle{$z\sim6$ protocluster}
\shortauthors{Toshikawa et al.}

\begin{document}

\title{Discovery of a protocluster at $z \sim 6$\altaffilmark{1}}
\author{Jun Toshikawa\altaffilmark{2}, Nobunari Kashikawa\altaffilmark{2,3}, Kazuaki Ota\altaffilmark{4}, 
    Tomoki Morokuma\altaffilmark{5}, Takatoshi Shibuya\altaffilmark{2}, \\Masao Hayashi\altaffilmark{3}, 
    Tohru Nagao\altaffilmark{4,6}, Linhua Jiang\altaffilmark{7}, Matthew A. Malkan\altaffilmark{8}, 
    Eiichi Egami\altaffilmark{7}, Kazuhiro Shimasaku\altaffilmark{9}, Kentaro Motohara\altaffilmark{5}, 
    and Yoshifumi Ishizaki\altaffilmark{2}}
\email{jun.toshikawa@nao.ac.jp}
\altaffiltext{1}{Based in part on data collected at Subaru Telescope, which is operated by the National 
    Astronomical Observatory of Japan.}
\altaffiltext{2}{Department of Astronomy, School of Science, Graduate University for Advanced Studies, 
    Mitaka, Tokyo 181-8588, Japan.}
\altaffiltext{3}{Optical and Infrared Astronomy Division, National Astronomical Observatory, 
    Mitaka, Tokyo 181-8588, Japan.}
\altaffiltext{4}{Department of Astronomy, Graduate School of Science, Kyoto University, Sakyo-ku, 
    Kyoto 606-8502, Japan.}
\altaffiltext{5}{Institute of Astronomy, University of Tokyo, Mitaka, Tokyo 181-0015, Japan.}
\altaffiltext{6}{The Hakubi Project, Kyoto University, Yoshida-Ushinomiya-cho, Sakyo-ku,
    Kyoto 606-8302, Japan.}
\altaffiltext{7}{Steward Observatory, University of Arizona, 933 North Chery Avenue, Tucson, AZ 85721.}
\altaffiltext{8}{Department of Physics and Astronomy, University of California, Los Angeles, CA 90095-1547.}
\altaffiltext{9}{Department of Astronomy, University of Tokyo, Hongo, Tokyo 113-0033, Japan.}

\begin{abstract}
We report the discovery of a protocluster at $z\sim6$ containing at least eight cluster member galaxies
with spectroscopic confirmations in the wide-field image of the Subaru Deep Field (SDF).
The overdensity of the protocluster is significant at the $6\sigma$ level, based on the surface number 
density of $i'$-dropout galaxies.
The overdense region covers $\sim6\arcmin\times6\arcmin \,(14\,\mathrm{Mpc}\times14\,\mathrm{Mpc}$ in 
comoving units at $z=6$), and includes 30 $i'$-dropout galaxies.
Follow-up spectroscopy revealed that 15 of these are real $z\sim6$ galaxies ($5.7<z<6.3$).
Eight of the 15 are clustering in a narrow redshift range ($\Delta z < 0.05$ centered at $z=6.01$),
corresponding to a seven-fold increase in number density over the average in redshift space.
We found no significant difference in the observed properties, such as Ly$\alpha$ luminosities and UV 
continuum magnitudes, between the eight protocluster members and the seven non-members.
The velocity dispersion of the eight protocluster members is $647\pm124\,\mathrm{km\,s^{-1}}$, which is about 
three times higher than that predicted by the standard cold dark matter model.
This discrepancy could be attributed to the distinguishing three-dimensional distribution
of the eight protocluster members.
We discuss two possible explanations for this discrepancy:
either the protocluster is already mature, with old galaxies at the center, or it is still immature and
composed of three subgroups merging to become a larger cluster.
In either case, this concentration of $z=6.01$ galaxies in the SDF may be one of the first sites
of formation of a galaxy cluster in the universe.
\end{abstract}


\section{INTRODUCTION}
Exploring the structure formation and evolutionary history of the early universe is an issue of 
strong current interest in astronomy.
In the cold dark matter (CDM) model, clusters of galaxies form in the densest peaks of dark matter in
the early universe.
These grow by merging and by accreting material from surrounding low-density regions \citep{springel05}.
Therefore, the growth of density perturbations measured by the comparisons of abundances and mass distributions
of present-day clusters with those at earlier times provides unique constraints on the $\Lambda$CDM
concordance model \citep[e.g.,][]{vikhlinin03,voit05,mortonson11}.
Theoretical models predict that galaxies lying inside these high-density regions may have formed
earlier and/or evolved more rapidly, compared with their surroundings
\citep[e.g.,][]{kauffmann95,benson01,thomas05,lucia06}.
Clusters of galaxies are, thus, the noteworthy site of dark matter structure formation and galaxy evolution.
In the local universe, cluster galaxies represent a distinct relationship in the color-magnitude diagram.
The ``red sequence'' in clusters is composed mainly of spheroidal and lenticular galaxies with old
stellar populations and high stellar masses \citep[e.g.,][]{visvanathan77,gladders00,lerchster11}.
Furthermore, brighter and massive galaxies are likely to lie the central region of a cluster.
When and how, in the history of the universe, were these distinct properties formed?
Highly evolved clusters have been intricately affected by both nature and nurture; thus, it is
difficult to sort out which factors are most important to cluster and galaxy evolution.
One promising line of study is to investigate directly the primitive properties of galaxies
belonging to the first clusters of galaxies in the early universe.
Protoclusters in the early universe would provide a great deal of information on the primordial conditions
of clusters at their birth.

Star-forming galaxies, such as Ly$\alpha$ emitters (LAEs) and Lyman break galaxies (LBGs), are almost the
only tracers that have been used to follow the evolution of large-scale structures at high redshifts.
We have very few clear examples of these star-forming galaxies strongly clustered beyond $z=3$ 
\citep[e.g.,][]{fevre96,giavalisco94,venemans05,kang09}.
Subsequent searches have been carried out at even higher redshifts
\citep[e.g.,][]{venemans02,venemans04,venemans07,overzier06,overzier08,capak11}.
Overdense regions of $i'$-dropout galaxies out to $z \sim 6$ have been found
\citep{malhotra05,stiavelli05,zheng06,kim09}; however, most of these potential protocluster members
have not been identified through spectroscopy.
Analysis based only on photometric data limits a detailed understanding of the physical processes, such
as three-dimensional structure, in the high-density environment.
Furthermore, sample could be contaminated by lower-redshift objects; it dilutes intrinsic properties of
protoclusters.

When cluster formation approaches its final stage, galaxy evolution is predominantly influenced by 
environmental effects.
Some differences between the properties of protocluster galaxies and field galaxies appear at 
$z\sim2\mathrm{-}3$: the masses of cluster galaxies are higher than those of field galaxies 
\citep{steidel05,kuiper10,hatch11}, and the fraction of galaxies harboring an active galactic nucleus (AGN) 
in protoclusters at $z\sim3$ is higher than that in the field \citep{lehmer09}.
These findings indicate that the first step in forming a cluster takes place at a much higher redshift.
In earlier epochs, at $z\sim4\mathrm{-}5$, \citet{overzier09b} found no significant difference in the 
stellar mass between protocluster galaxies and field galaxies.
Additionally, \citet{overzier09b} showed that the total stellar mass of protocluster galaxies at $z\sim4$
is much smaller than that of massive cluster galaxies at $z\sim1$ which are expected to have formed at $z\ga4$.
This may suggest that there are more massive structures at $z>4$, which will become rich cluster
in the local universe.
It is important to investigate protoclusters at even higher redshifts directly, to understand how the 
clusters are formed and how galaxy evolution depends on the environment.
Investigating protoclusters during the epoch of reionization may also be important for understanding 
the early history of the universe, because the reionization process and thermal histories in overdense 
regions are expected to be remarkably different from those of average density, due to the enhanced number
of ionizing sources and higher radiative feedback \citep[e.g.,][]{mcquinn07,iliev08}.

Most of these protoclusters have been discovered in regions centered on radio galaxies (RGs) or quasars (QSOs)
\citep{miley08}.
RGs and QSOs have been used as potentially useful probes of large-scale structure, as signposts of possible 
regions of galaxy overdensity. 
However, strong radiation from RGs or QSOs may provide contradictory feedback that suppresses nearby galaxy 
formation, especially of low-mass galaxies \citep{barkana99}.
\citet{kashikawa07} found a ring-like structure of LAEs around a luminous QSO at $z\sim5$, possible 
evidence of negative feedback on star-formation activity in galaxies neighboring QSOs.
Due to their low number density, it is difficult to discover protoclusters without RGs or QSOs 
at $z>3$.
A few protoclusters or large-scale structures have been discovered serendipitously in random fields at high 
redshift \citep[e.g.,][]{steidel98,shimasaku03,ouchi05,lemaux09}, suggesting that early massive structures do
not always host RGs/QSOs.

In this paper, we present the most distant protocluster candidate known, at $z=6$, hosting at least eight 
spectroscopically confirmed members.
This protocluster was found from an examination of data taken with a wide field-of-view (FOV) camera
mounted at the prime focus of the Subaru telescope and aimed at the Subaru Deep Field (SDF;
RA=$13^\mathrm{h}24^\mathrm{m}38\fs9$, DEC=$+27\arcdeg29\arcmin25\farcs9$ [J2000]; \citet{kashikawa04}).
The prime focus camera, ``SuprimeCam'' \citep{miyazaki02}, has an FOV of $34' \times 27'$, and is a powerful
instrument for discovering similar rare objects.
This paper is organized as follows.
\S2 describes the imaging data used in this study and our selection of $z\sim6$ galaxies.
In \S3, we determine the significance of the surface number density of the $z\sim6$ galaxies 
in the overdense region.
In \S4, we describe our follow-up spectroscopic observations.
In \S5, we discuss the properties and the structure of the protocluster.
The conclusions are given in \S6.
We assume the following cosmological parameters: $\Omega_\mathrm{M}=0.3, \Omega_\Lambda=0.7, 
\mathrm{H}_0=70 \mathrm{\,km\,s^{-1}\,Mpc^{-1}}$, which yield an age of the universe of $910\,\mathrm{Myr}$
and a spatial scale of $40\,\mathrm{kpc\,arcsec^{-1}}$ in comoving units at $z=6$.
Unless otherwise noted, we used comoving units throughout.
Magnitudes are given in the AB system, and we used a $2\arcsec$ aperture.

\section{SAMPLE SELECTION}
\subsection{Photometric Data}
We used the SDF public data, with limiting magnitudes of $B=28.45$, $V=27.74$, $R=27.80$, $i'=27.43$, 
and $z'=26.62$ (2{\arcsec} aperture, 3$\sigma$).
We have also obtained new, deeper $R$-, $i'$-, and $z'$-band images.
These images were constructed by stacking all the data taken from 2001 to 2008 in the course of a study
of distant supernovae \citep{poznanski07,graur11}, containing almost 30 hours worth of integration
time in total.
The $3\sigma$ limiting magnitudes of these new deep images are $28.35$, $27.72$,
and $27.09$ at $R$-, $i'$-, and $z'$-bands, respectively. 
These are about 0.5 mag deeper than the SDF public data.
We selected $i'$-dropout objects using these deep $R$-, $i'$-, $z'$-band images 
and the public $B$- and $V$-band images.
All five images were convolved to a common seeing size of $0\farcs98$.
We obtained $J$-band imaging of the SDF with the WFCAM on UKIRT \citep{casali07} in March, April, and
July 2010 (Hayashi et al., in prep).
The SDF was entirely covered by a mosaic $J$-band image, although the depth was not uniform from field to field.
We supplemented our data with this $J$-band imaging, the limiting magnitude of which was $\ga2$ mag
shallower than the $z'$-band image. 
The seeing size is $1\farcs1$.
This allowed us to reject apparent contamination by M/L/T dwarfs.
The pixel scales of all six images are $0\farcs20$, and the details of the image properties are summarized in
Table \ref{photo}.

\begin{deluxetable*}{ccccccc}
\tabletypesize{\scriptsize}
\tablecaption{Photometric Data \label{photo}}
\tablewidth{0pt}
\tablehead{ & $B$ & $V$ & $R$ & $i'$ & $z'$ & $J$\tablenotemark{a}}
\startdata
$m_{lim,3\sigma}$\tablenotemark{b} & 28.57 & 27.85 & 28.35 & 27.72 & 27.09 & 23.30-24.80 \\
integration time(hour) & 9.9 & 5.7 & 27 & 27.7 & 30.9 & 1.1-10
\enddata
\tablenotetext{a}{$J$-band image is not uniform over the field, due to mosaicing four regions.
    The protocluster region is located in the deepest portion of the $J$-band image.}
\tablenotetext{b}{$3\sigma$ limiting magnitude in a $2\arcsec$ aperture}
\end{deluxetable*}

We performed object detection and photometry by running SExtractor (ver. 2.5.0) \citep{bertin96} on the images.
Object detections were made in the $z'$-band.
Then, the magnitudes, and several other photometric parameters, were measured in the other bands at exactly 
the same positions and with the same apertures as in the detection-band image.
This task used the SExtractor ``double image mode.''
We marked objects that had five connected pixels with signal that was $2\sigma$ above the sky 
background RMS noise. 
Photometric measurements were made at the $2\sigma$ level.
Objects detected in regions with low S/N ratio, near the frame edges or near saturated 
pixels around bright stars, were removed from the catalog.
The remaining effective area of analysis was 876 $\mathrm{arcmin^2}$.
Finally, $\sim102000$ objects were detected down to $z'=27.09$ ($3\sigma$ limiting magnitude).
To estimate the detection completeness of the $z'$-band image, we used the IRAF task {\sf mkobjects} to 
create artificial objects on the original image.
Artificial objects were created with the same FWHM ($0\farcs98$) as real images, and were randomly distributed.
To avoid blending artificial objects with real objects, we avoided positions close to the real objects
with distances shorter than 1.5 times the FWHM of the real objects.
We extracted the artificial objects using SExtractor with the same parameter set.
We generated 3000 artificial objects in the 23 to 29 magnitude range, and repeated this procedure 20 times.
The detection completeness was more than 90\% at $z'=25$ and 70\% at $z'=27.1$ ($3\sigma$ limiting magnitude).

\subsection{Selection Criteria of $z\sim6$ Galaxies}
\begin{figure}
\epsscale{1.17}
\plotone{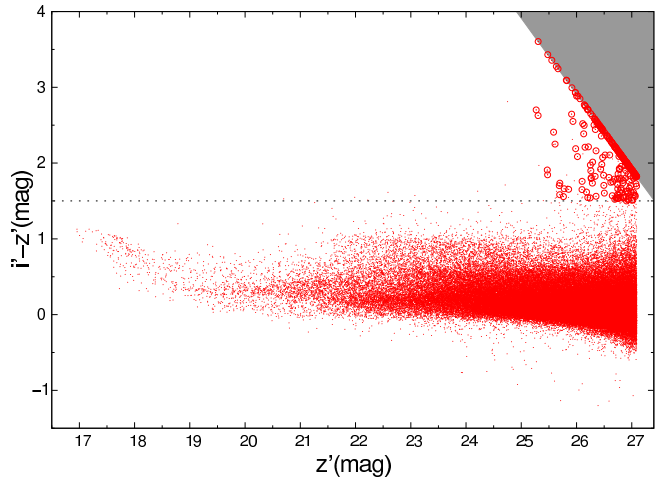}
\caption{A color-magnitude plot of detected objects (dots) brighter than the $3\sigma$ limiting magnitude
    ($z'=27.09$) in $z'$-band.
    The horizontal dotted line shows the color criterion $i'-z'=1.5$ for $z\sim6$ galaxy candidates, 
    and the gray region shows $i'$-band magnitudes fainter than $1\sigma$ limiting magnitude ($i'=28.91$).
    The open circles are the 258 $z\sim6$ galaxy candidates. \label{col_mag}}
\end{figure}

Figure \ref{col_mag} presents a color-magnitude diagram of the $z'$-band detected objects.
We selected $z\sim6$ galaxy candidates using the following criteria: 1) $i'-z' \ge 1.50$
down to $z'<27.09$ (The $1\sigma$ limiting magnitude of $i'$-band was used as $i'-z'$ color limit, if
objects were not detected in $i'$-band image), 2) null detection ($< 2 \sigma$) in $B$- and $V$-bands, 3)
$z'-J \le 0.80$ (this criterion was imposed only for objects detected in $J$-band.), and 4) $z' \ge 25.0$.
The $1^\mathrm{st}$ color criterion picked up galaxies with Lyman breaks at $5.6 \la z \la 6.9$, based on
calculations from the population synthesis model code GALAXEV \citep{BC03} and intergalactic absorption
\citep{madau95}.
In GALAXEV, we simulated a large variety of galaxy spectral energy distributions (SEDs) using the Padova 1994
simple stellar population model.
We assumed a \citet{salpeter55} initial mass function with lower and upper mass 
cutoffs $m_L=0.1\,\mathrm{M_\sun}$ and $m_U=100\,\mathrm{M_\sun}$, two metallicities (0.2 and 0.02 
$\mathrm{Z_\sun}$), and instantaneous burst star formation history.
We extracted model spectra with ages between $10\,\mathrm{Myr}$ and $740\,\mathrm{Myr}$ and applied the 
reddening law of \citet{calzetti00} with $E(B-V)$ of 0.00, 0.01 and 0.10.
The $2^\mathrm{nd}$ criterion was required because the flux at the wavelength blueward of
$\mathrm{Ly_{limit}}$ for $z\sim6$ galaxies should be absorbed almost completely by the intergalactic medium.
We did not require null detections in either the $R$- or $i'$-band because the continuum blueward of Ly$\alpha$
may not be totally extincted at $z\sim6$ and because the Ly$\beta$ and Ly$\gamma$ emissions, if they are present
in these galaxies, can fall in these bands.
However, it is interesting to note that the selected objects have fainter $R$ magnitudes than the $i'$
magnitude, or are not detected in the $R$-band.
The $3^\mathrm{rd}$ criterion was required because the spectrum of $z\sim6$ LBG redward of Ly$\alpha$ 
should be blue: UV-continuum slopes, $\beta$ ($f_\lambda\propto\lambda^\beta$), are typically
$\beta\la-2$ \citep[e.g.,][]{bouwens09}.
As mentioned in \S2.1, our $J$-band image was $>2\;\mathrm{mag}$ shallower than the $z'$-band image;
thus, almost no $z\sim6$ galaxies should be detected in the $J$-band image.
The $3^\mathrm{rd}$ criterion was imposed only for objects detected in $J$-band.
Only nine objects of 267 $i'$-dropout objects were detected in the $J$-band image, 
and eight objects with $z'-J \ge 0.8$ were removed from the sample.
The $4^\mathrm{th}$ criterion was also required to keep the contamination rate low.
One object, which did not satisfy the $4^\mathrm{th}$ criterion, was removed from the sample.
Finally, we obtained 258 $z\sim6$ galaxy candidates.

We evaluated the contamination rate for these color-selection criteria.
The sources of the majority of the contamination were M/L/T dwarfs and $z\sim1.3$ old elliptical galaxies,
the latter being able to satisfy the $1^\mathrm{st}$ criterion ($i'-z'\ge1.50$) due to the 4000{\AA}
Balmer break.
We simulated old galaxy SEDs at $z\sim1.3$ using the GALAXEV, assuming two relatively high metallicities
($\mathrm{Z_\sun}$ and $2.5\mathrm{Z_\sun}$), and extracted model spectra with age of
$0.5 \mathrm{-} 5\;\mathrm{Gyr}$, applying reddening with $E(B-V)$ of 0.5, 1.0, and 1.5.
Based on these simulations, high amount of dust ($E(B-V)\ga1.5$) is needed to meet the
$1^\mathrm{st}$ color criterion.
According to the Extremely Red Objects (EROs) catalog by \citet{miyazaki03}, 96\% of EROs are $i'-z'<1.5$ and
no EROs have $E(B-V)\ga1.5$ at $z\sim1.3$.
Additionally, \citet{malhotra05} show that $i'$-dropout objects with $i'-z'>1.3$ do not include any EROs
based on their spectroscopy.
Furthermore, the $2^\mathrm{nd}$ and $3^\mathrm{rd}$ criteria are supplementarily used in order to
discriminate low-$z$ galaxies from $z\sim6$ galaxies.
Hence, we ignore the possible contamination by low-$z$ galaxies in the subsequent analyses and discussions.
Secondly, we consider contamination by M/L/T dwarfs.
According to \citet{hawley02}, M/L/T dwarfs that meet the $1^\mathrm{st}$ criterion should have a very
red color ($z'-J\ga2$).
Thus, M/L/T dwarfs can be, in principle, detected in the $J$-band image such that they would be removed
from our $z\sim6$ galaxy candidate sample with the $3^\mathrm{rd}$ criterion.
However, since the $J$-band limiting magnitude is $>2$ mag shallower than that of the $z'$-band, in practice
the $3^\mathrm{rd}$ criterion is effective to remove only the brightest or very reddest M/L/T dwarfs from
our high-redshift sample.
The $4^\mathrm{th}$ criterion was also imposed to restrain the contamination rate by M/L/T dwarfs because
the contamination rate is high at brighter than $z'\sim25$.
From the star count model developed by \citet{nakajima00}, the contamination rate at $z'>25$ is only about
6\% at the galactic coordinates corresponding to our survey area.
Therefore, we assumed a contamination rate in our $z\sim6$ galaxy selection of up to a few percent,
mainly consisting of contamination from fainter or bluer M/L/T dwarfs.

\begin{figure}
\epsscale{1.25}
\plotone{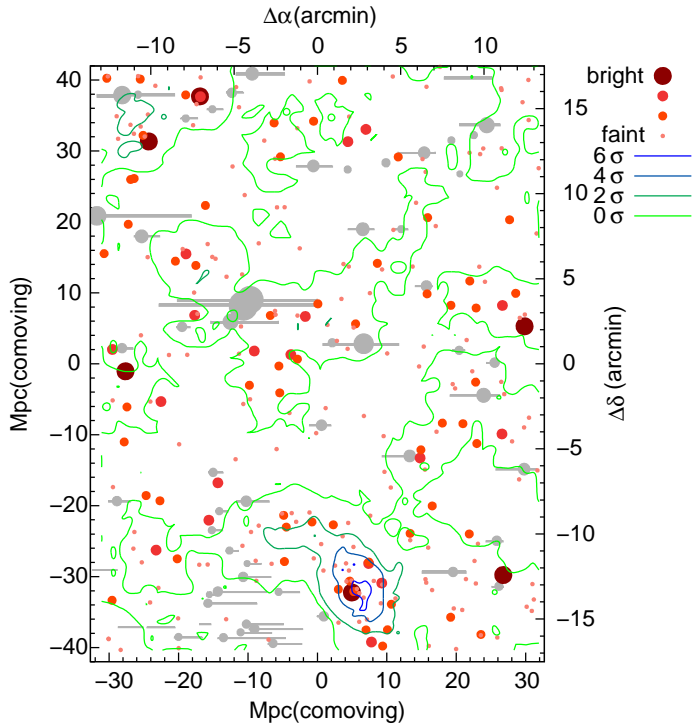}
\caption{The sky distribution of 258 $z\sim6$ galaxy candidates, with surface number density contours.
    The $z\sim6$ galaxy candidates are represented by filled circles whose size is proportional to the 
    $z'$-band magnitudes (huge: $25.0 \le z'<25.5$, large: $25.5\le z'<26.0$, medium: $26.0\le z'<26.5$, 
    small: $26.5 \le z' \le 27.1$).
    The lines correspond to contours of surface overdense significance from $6\sigma$ to $0\sigma$
    with a step of $2\sigma$.
    North is up and east is to the left.
    The comoving scale projected to $z=6$ is also shown along the axes.
    Gray regions are masked regions.
    The overdense region can be clearly seen at the southern edge of the plot. \label{cntr}}
\end{figure}

In addition to these intrinsic contaminates, photometric noise may scatter lower-redshift objects to
satisfy our sample selection criteria.
We performed the following simple simulation, the same as is performed in \citet{wilkins11a} and
\citet{bouwens11a}, to estimate the contamination rate due to the photometric noise.
We first randomly choose brighter objects at $22.0<z'<22.5$, dim these bright objects so as to match
the magnitude distribution to those of our $i'$-dropout at $25.0<z'<27.09$ objects by scaling the flux,
then distribute these artificially dimmed objects on the original image as we did in \S2.1.
In this simulation, we simply assumed that the intrinsic color distribution of faint objects is almost
the same as that of brighter objects whose photometric noise should be negligible.
The bright sources are chosen from only slightly brighter magnitude ranges, in which they have higher S/N but
are expected to have similar color distribution with our $i'$-dropout sample.
We extracted the artificial objects using SExtractor and imposed our color criteria of the $i'$-dropout
objects.
The number of artificial objects in a given $z'$-magnitude interval was chosen to be the same as the
observed number of object in the same $z'$-magnitude interval.
We finally found that 39 artificial objects that met our color criteria, {\it i.e.}, 15\% of 258
$i'$-dropout objects may be contaminated due to the photometric noise.
It should be noted that these contaminants appeared almost randomly over the image in the repeated simulation,
although the depth of $J$-band image is not uniform over the field.
This suggests that the contamination rate is almost homogeneous over the survey field and did not change the
overdensity significance estimated in the next section.

\begin{deluxetable}{ccccc}
\tabletypesize{\scriptsize}
\tablecaption{Photometric Properties of the 15 Spectroscopic Confirmed Galaxies \label{phot_cat}}
\tablewidth{0pt}
\tablehead{\colhead{ID} & \colhead{RA} & \colhead{DEC} & \colhead{$z'$} & \colhead{$i'-z'$} \\
            \colhead{} & \colhead{(J2000)} & \colhead{(J2000)} & \colhead{(mag)} & \colhead{(mag)}}
\startdata
1\tablenotemark{a} &13:24:31.8 &+27:18:44.2 &$25.90\pm0.04$ &1.48 \\ %MOS8-obj4
2 &13:24:18.4 &+27:16:32.6 &$25.69\pm0.03$ &1.70 \\ %sdf13-slit16
3 &13:24:25.2 &+27:16:12.2 &$27.02\pm0.09$ &$>1.89$ \\ %MOS7-obj19
4 &13:24:30.2 &+27:14:13.5 &$26.81\pm0.08$ &1.96 \\ %MOS7-obj15
5\tablenotemark{b} &13:24:26.0 &+27:16:03.0 &$26.50\pm0.06$ &1.71 \\ %MOS7-obj9
6 &13:24:21.3 &+27:13:04.8 &$25.91\pm0.04$ &2.65 \\ %MOS7-obj3
7 &13:24:29.0 &+27:19:18.0 &$26.50\pm0.06$ &1.60 \\ %MOS2-obj11
8\tablenotemark{a} &13:24:28.1 &+27:19:32.7 &$26.10\pm0.04$ &1.54 \\ %MOS8-obj5
9 &13:24:26.5 &+27:15:59.7 &$25.47\pm0.03$ &1.91 \\ %sdf6-slit18
10\tablenotemark{a} &13:24:31.5 &+27:15:08.8 &$25.91\pm0.06$ &1.95 \\ %sdf6-slit71
11 &13:24:26.1 &+27:18:40.5 &$26.59\pm0.06$ &$>2.32$ \\ %sdf27-slit11
12 &13:24:31.6 &+27:19:58.2 &$26.47\pm0.06$ &2.07 \\ %MOS8-obj9
13 &13:24:44.3 &+27:19:50.0 &$26.43\pm0.06$ &$>2.49$ \\ %MOS8-obj8
14\tablenotemark{c} &13:24:20.6 &+27:16:40.5 &$27.01\pm0.09$ &$>1.90$ \\ %MOS7-obj18
15\tablenotemark{a} &13:24:32.6 &+27:19:04.0 &$25.54\pm0.03$ &1.47 %MOS8-obj3
\enddata
\tablenotetext{a}{These do not satisfy all of our selection criteria.}
\tablenotetext{b}{The same object as No. 6 in Table 1 of \citet{jiang11}.}
\tablenotetext{c}{The same object as No. 7 in Table 1 of \citet{jiang11}.}
\end{deluxetable}

\section{SKY DISTRIBUTION AND OVERDENSITY SIGNIFICANCE}
Figure \ref{cntr} shows the sky distribution of the 258 $z\sim6$ galaxy candidates.
We found an apparently overdense region, centered in the southern part of the field.
To determine the overdensity significance quantitatively, we estimated the local surface number density
by counting $z\sim6$ galaxy candidates within a fixed aperture of 2\farcm1 radius, corresponding to
$5\,\mathrm{Mpc}$ at $z=6$.
These apertures were distributed over the SDF in a grid pattern at intervals of 20 arcsec.
We assumed that the local surface number density in masked regions was the same as the mean surface 
number density.
The mean and the dispersion, $\sigma$, of the number of $z\sim6$ galaxy candidates in a circle was 
found to be 4.0 and 2.6 per cell, respectively.
Based of these values, we also plotted surface number density contours in
Figure \ref{cntr}. 
We found that our region of interest appeared overdense at the $6\sigma$ significance level beyond
the mean surface number density at the peak.
We designated an ``overdense region'' $6'\times6'\,(14\times14\,\mathrm{Mpc^2})$ centered on 
the peak of overdensity.
This encompasses a region with an overdensity of more than $3\sigma$ significance.
This overdense region included 30 $z\sim6$ galaxy candidates.
The size of the overdense region was almost the same as that of the protoclusters at 
$z=2 \mathrm{-} 5$ \citep{venemans07}.
\citet{ouchi05} found two protoclusters whose sizes were $\sim7$ Mpc diameter at $z=5.7$ in the
Subaru/XMM-Newton Deep Survey.
\citet{kang09} also found an overdense region with $\sim10\,\mathrm{Mpc}$ diameter at $z=3.7$ in Chandra 
Deep Field-South.
From these comparisons, the size of our overdense region is almost consistent with previous 
observational results of protoclusters at lower-$z$.

\begin{figure}
\epsscale{1.19}
\plotone{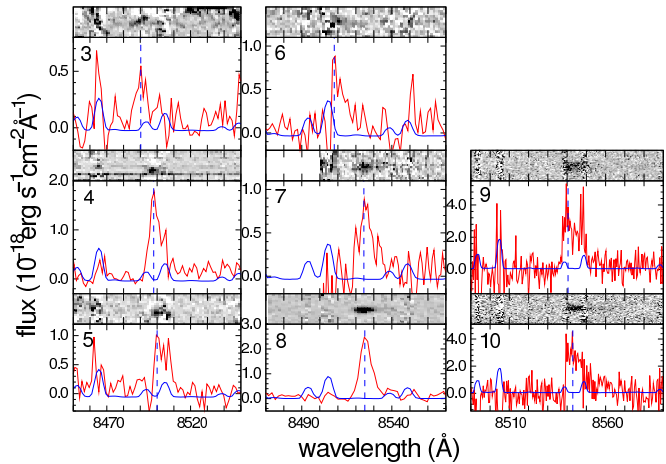}
\caption{Spectra of eight protocluster member galaxies.
    The vertical dashed lines indicate the peak of Ly$\alpha$ emission line.
    Blue solid lines represent sky lines.
    The numbers at the upper left corner are the IDs (Column 1 of Table 3). \label{spec}}
\end{figure}

We also compared our overdense region with theoretical models.
\citet{overzier09a} constructed a very large ($\sim4\arcdeg\times4\arcdeg$) mock redshift survey
of star-forming galaxies at $z\sim6$, using the Millennium Run dark matter simulation \citep{springel05} 
and semi-analytic prescriptions of galaxy formation.
In this mock survey, the limiting magnitude of $z'$-band was $26.5\mathrm{\,mag}$.
They predicted a typical protocluster diameter of not more than $\sim 25$ Mpc at $z\sim6$.
The size of our overdense region ($\la14\mathrm{\,Mpc}$) is consistent with this theoretical result.
Furthermore, they calculated the probabilities of finding overdense regions and the fraction of 
$i'$-dropout objects marked as protocluster member galaxies.
Applying the $i'$-dropout selection criteria of \citet{overzier09a} to our catalog, the probability of 
finding a region with an overdensity significance higher than ours within a 
$3.4\times3.4\,\mathrm{arcmin^2}$ (the FOV of HST/ACS) is only about 0.3\%; the local 
surface number density was estimated following the way of \citet{overzier09a}.
Furthermore, it is estimated from these results that 50\% of $z\sim6$ photometric candidates in
this overdense region belong to the protocluster, though the scatter is considerable ($1\sigma$ range
is between 0.3 and 0.7).
The overdense region contains fainter galaxies at $26.5<z'<27.1$; the probability of these being real
protocluster members is not predicted.
The discovery of an overdense region with such a high statistical significance would be extremely rare,
and half of all galaxies in the region are expected to be protocluster galaxies.
Therefore, there is a high probability that the overdense region includes a protocluster at $z\sim6$.

\section{SPECTROSCOPIC CONFIRMATION}
Our discovery of an overdense region might be attributed to a mere chance of alignment,
because the dropout technique samples a broader range of redshifts.
It might also be an incidental result of highly clustered contaminating populations.
To investigate this, we carried out follow-up spectroscopy on the overdense region, using FOCAS
\citep{kashikawa02} on the Subaru telescope in Multi-Object Spectroscopy (MOS) mode.
The details of this observation will appear elsewhere \citep{shibuya11}.
The FOCAS observations were made with a $555\,\mathrm{mm^{-1}}$ grating and an O58 order-cut filter,
giving spectral coverage of 7500-10450{\AA} with a resolution of $0.74\mathrm{\,\AA\,pix^{-1}}$.
The $0\farcs8$-wide slit gave a spectroscopic resolution of 5.7{\AA} (R$\sim$1500).
We targeted 27 $z\sim6$ galaxy candidates in the overdense region.
Additionally, we allocated slits to eight objects that did not meet our photometric criteria perfectly,
i.e., $i'-z'\ge1.40$, $z'<25.0$, or $B$-band magnitude slightly brighter than 
the $2\sigma$ limiting magnitude, to evaluate our selection criteria.
The data were obtained on 2010 March 19 and 20.
Sky conditions were good with a seeing size of $0\farcs5$-$1\farcs0$.
We obtained eight exposures, for a total integration time of four hours.
Between exposures, the telescope was dithered along the slit to enable more accurate sky subtraction.
Long slit exposures of standard star HZ 44 were used for flux calibration.
The data were reduced in a standard manner with IRAF.

\begin{figure}
\epsscale{1.19}
\plotone{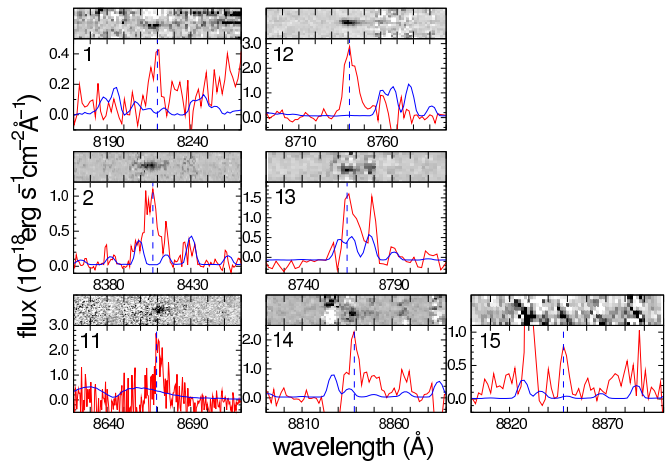}
\caption{Same as Figure \ref{spec}, but for the seven non-member galaxies. \label{nonspec}}
\end{figure}

We carefully discriminated real emission lines from sky lines by examining 2D and 1D spectra, 
and detected single emission lines from 10 objects of 35 targets.
Our spectroscopic observations were carried out with $R\sim1500$, which is a high enough resolving
power, in principle, to distinguish single Ly$\alpha$ emission from $[$O{\sc ii}$]$ doublet
($\Delta \lambda = 6.3\,${\AA} at $\sim8500\,${\AA}), although it is difficult in practice to discriminate
between the features if the emission is very faint.
Rather than relying on the resolving power of our spectroscopy, we use the emission line profile as
the primary diagnostic for distinguishing Ly$\alpha$ at high redshift from $[$O{\sc ii}$]$ emission.
We calculated the weighted skewness, $S_w$ \citep{kashikawa06}, which is a good indicator of the 
Ly$\alpha$ emission-line asymmetry.
The results are listed in Table 3.
We found that most of the detected emission lines had $S_w>3$, indicating that they are certainly
Ly$\alpha$ emission lines at $z\sim6$; however, some emission lines did not meet the Ly$\alpha$ criterion.
These line profiles may be affected by strong sky lines in this wavelength range, or low S/N ratio 
data could prevent accurate determination of the skewness.
As mentioned in \S2.1, it is unlikely that $[$O{\sc ii}$]$ emitters at $z\sim1.3$ satisfy
the photometric selection criterion of $i'-z'\ge1.50$.
Although we cannot completely deny the possibility that objects with $S_w<3$ were $[$O{\sc ii}$]$ 
emission lines, the single emission lines from our $i'$-dropout sample were plausibly identified as 
Ly$\alpha$ emission at $z\sim6$.
We hereafter regard the single emission line detected from our $i'$-dropout sample as Ly$\alpha$ emission.

\begin{figure}
\epsscale{1.05}
\plotone{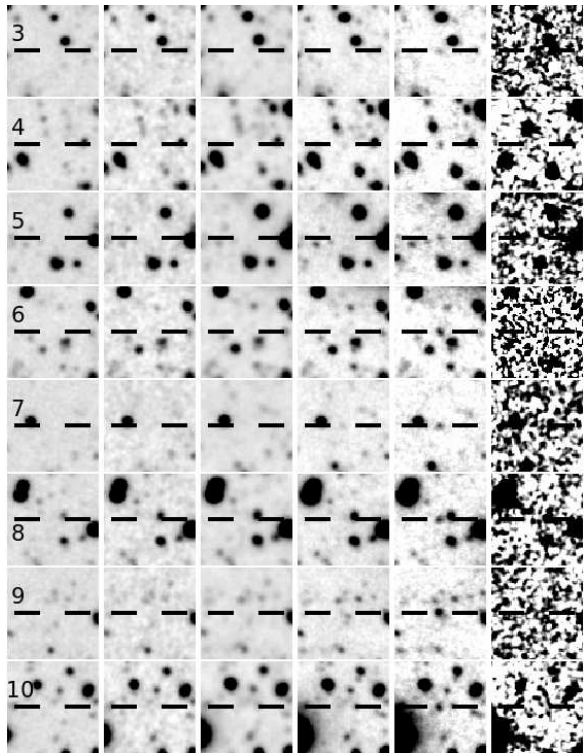}
\caption{Thumbnail images ($\sim15\arcsec\times15\arcsec$) of eight protocluster member
    galaxies in six passbands.
    From left to right, the $B$-, $V$-, $R$-, $i'$-, $z'$-, and $J$-band images are shown.
    North is up and east is to the left.
    The numbers at the upper left corner are the IDs (Column 1 of Table 3). \label{stamp}}
\end{figure}

We detected Ly$\alpha$ emissions from seven objects of 27 targeted $z\sim6$ galaxy candidates that met
all of our photometric criteria.
We did not detect any signal from the remaining 20 objects.
No apparent contaminations, such as M/L/T dwarfs or low-$z$ galaxies, were detected.
We allocated slits to eight candidates that did not meet our photometric criteria: three of these
were identified as Ly$\alpha$ emitters at $z\sim6$, three were apparent stars.
We did not detect any signal from the remaining two objects.

We also obtained five Ly$\alpha$ emission lines at $z\sim6$ in the overdense region 
from previous observations, carried out with Subaru/FOCAS or Keck/DEIMOS \citep{faber03}.
These spectra were obtained through previous MOS observations on several different projects 
\citep{nagao07,ota08,kashikawa11,jiang11,shibuya11}.
Our $z\sim6$ galaxy photometric candidate sample includes four of these five galaxies.
The remaining $z\sim6$ galaxy was so close to a bright star that we could not detect it.

\begin{figure}
\epsscale{1.05}
\plotone{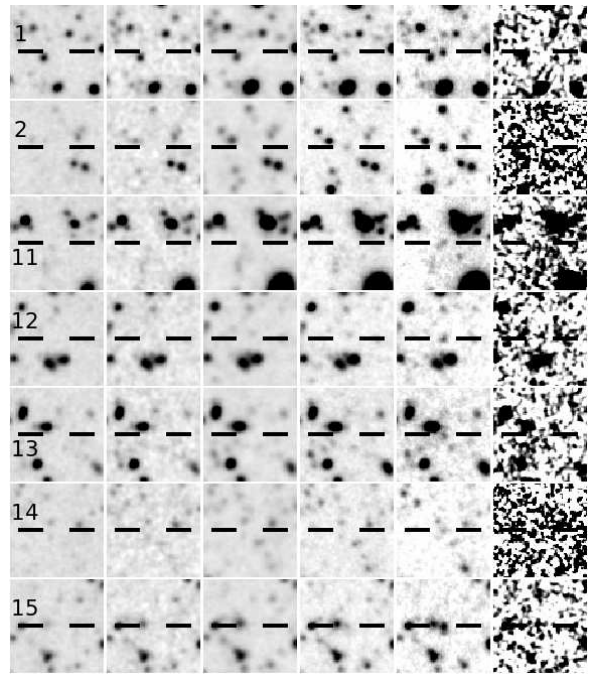}
\caption{Same as Figure \ref{stamp}, but for the seven non-member galaxies. \label{nonstamp}}
\end{figure}

In sum, we spectroscopically confirmed 15 $z \sim 6$ galaxies, including four serendipitous objects, 
in the overdense region.
The details of all 15 $z\sim6$ galaxies are provided in Table \ref{phot_cat} and 3.
The spectra and thumbnail images of all 15 $z\sim6$ galaxies are shown in Figure \ref{spec}, 
\ref{nonspec} and \ref{stamp}, \ref{nonstamp}, respectively.
All the $z\sim6$ galaxies with measured redshifts are identified by their Ly$\alpha$ emissions.

The spectroscopically undetected targets were expected to have a Ly$\alpha$ flux that is too low to be
detected (or non-existent).
Our sample was selected based on the UV continuum flux using broad-band images.
Therefore, these were not necessarily required to have Ly$\alpha$ emission.
As noted in \S2.1, our $z'$-band image, with a 30-hour-long integration time, was stacked from all
the data taken for 8 years.
Therefore, there is little chance that they are variable objects, such as supernovae or active galactic
nuclei that incidentally increased their luminosity during the $z'$-band observation .
The misalignment of slits is also quite unlikely because we used the MOS mode with enough alignment stars,
and we actually succeed in detecting signals in some of slits in the same MOS mask. 
The detection rate of Ly$\alpha$ emission of $i'$-dropout candidates is 75\% at $z'\le26$, 38\% at
$26<z'\le26.5$, and 22\% at $z'>26.5$.

\begin{figure}
\epsscale{1.23}
\plotone{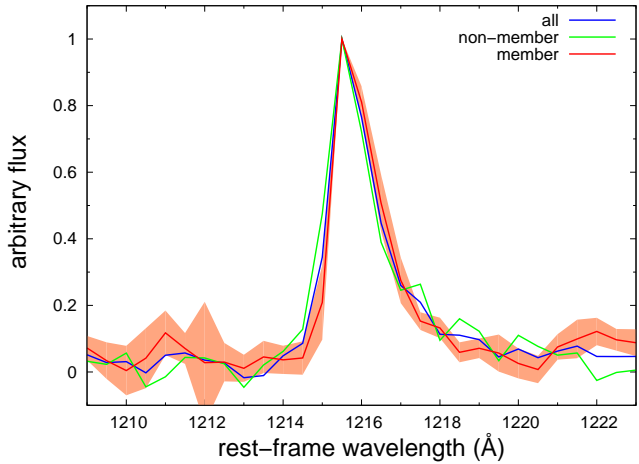}
\caption{The composite spectra of spectroscopically confirmed galaxies.
    The blue, red, and green lines indicate that of all 15 galaxies,
    eight protocluster member galaxies, and seven protocluster non-member galaxies, respectively.
    The red-shaded region shows the $1\sigma$ variance on the composite spectrum of member galaxies.
    \label{comp}}
\end{figure}

We made a composite spectrum of all 15 Ly$\alpha$ emission lines.
Each galaxy spectrum was shifted to its rest-frame and normalized using the flux at peak.
Then, we created a weighted average of all spectra in which the weight was determined from the S/N ratio.
The resulting composite spectrum is shown in Figure \ref{comp}, clearly revealing an asymmetric profile,
a characteristic feature of Ly$\alpha$ emission lines at high redshift.
The weighted skewness was found to be $S_w=8.76\pm1.05$.

The redshifts of eight galaxies (ID=3-10 in Table 3) of 15 LAEs in the
overdense region lie between $z=5.984$ and $z=6.027$, apparently concentrating at $z=6.01$ ($\Delta z < 0.05$).
Figure \ref{hist} shows the redshift distribution of 15 galaxies.
The redshift concentration at $z=6.01$ is about seven times higher than the number density expected 
from a uniform distribution.
Moreover, it should be noted that all eight galaxies clustered in redshift have $S_w>3$, suggesting that
these are plausible Ly$\alpha$ emitters in a protocluster at $z\sim6$.
Figure \ref{cluster} shows the three-dimensional distribution of these eight galaxies.
These eight galaxies are strongly clustered, with separations less than $16\,\mathrm{Mpc}$.
Thus, this region is almost certain to be a galaxy protocluster at $z=6.01$.
If true, this protocluster would be the highest redshift large scale structure probed by the 
combination of the $i'$-dropout technique and the unique wide-field imaging capability of SuprimeCam.

\section{DISCUSSION}
\subsection{Uncertainties in the Overdense Significance}
We applied a color threshold $i'-z'\ge1.50$ to select $i'$-dropout objects, although the critical 
threshold does not perfectly eliminate other contaminations, and photometric errors dilute the 
discrimination.
We evaluated the effect of the critical color threshold on the significance of the overdense region
by changing the threshold of the $1^\mathrm{st}$ criterion from 1.5 to 1.3, 1.4, 1.6 and 1.7 in the
sample selection.
We re-estimated the significance of the overdense region relative to spatially uniform distribution
with the same procedure as in \S3.
The overdense region was always uniquely identified as the highest-density region in the SDF,
and the peak significance fluctuated between $5\sigma$ and $7\sigma$, indicating that the significance 
was statistically robust.
As noted in \S2.2, the contamination from M/L/T dwarfs could be varied from field to field due to non-uniform
$J$-band depth over the field.
However, since the overdensity region is mostly located in the deepest portion of the mosaic $J$-band image,
the overdensity significance should not be incidentally enhanced by the contamination.

The spectroscopic identification of Ly$\alpha$ emitters has some uncertainty.
In \S4, we regarded all single emission lines detected from our $i'$-dropout sample as Ly$\alpha$
emission, although we found some emissions with asymmetries which were not high enough to be 
conclusively identified as Ly$\alpha$.
The objects labeled as ID=1 and 15 in Table 3, have $S_w<3$.
These might be $[$O{\sc ii}$]$ emitters.
It should be noted that all eight protocluster galaxies have $S_w>3$.
Even if the objects with $S_w<3$ were deleted from the $z\sim6$ galaxy, the concentration of the eight
galaxies still remains.
Furthermore, the detection limit of Ly$\alpha$ emission is not, generally, a uniform function of wavelength.
This is due to the effect of strong night sky lines in certain spectral window, which can overwhelm the faint
Ly$\alpha$ emission we are attempting to detect.
The fact that all of our protocluster members are detected within $8520\pm30\,\mathrm{\AA}$, a window not free
of bright airglow lines, further bolsters our confidence that the redshift clustering of our targets has an 
astrophysical origin and is not a consequence of selection effects resulting from night sky emission.

\begin{figure}
\epsscale{1.18}
\plotone{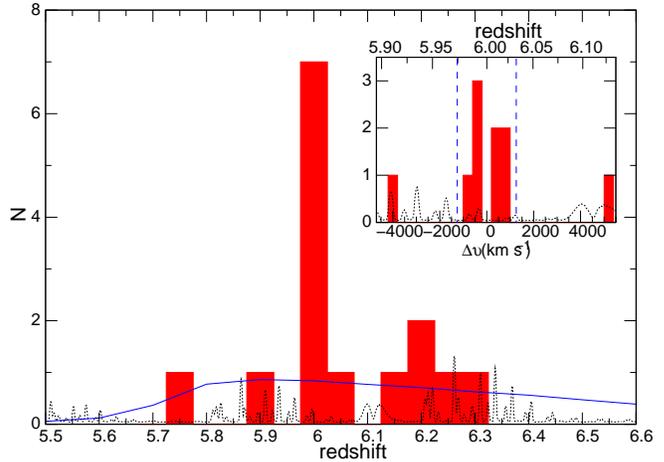}
\caption{The redshift distribution of 15 spectroscopically confirmed galaxies.
    The bin size is $\Delta z = 0.05$.
    The solid (blue) line shows the selection function of our $i'$-dropout selection assuming a uniform 
    distribution normalized to the total number of confirmed emitters.
    The inset is a close-up of the protocluster redshift range, with a bin size of $400\,\mathrm{km\,s^{-1}}$.
    The horizontal axis, $\Delta v$, represents radial velocity.
    The eight protocluster members are distributed between the vertical dashed (blue) lines.
    The dotted lines show the location of sky lines if redshifts are determined by Ly$\alpha$ emission.
    \label{hist}}
\end{figure}

\begin{figure*}
\epsscale{0.9}
\plotone{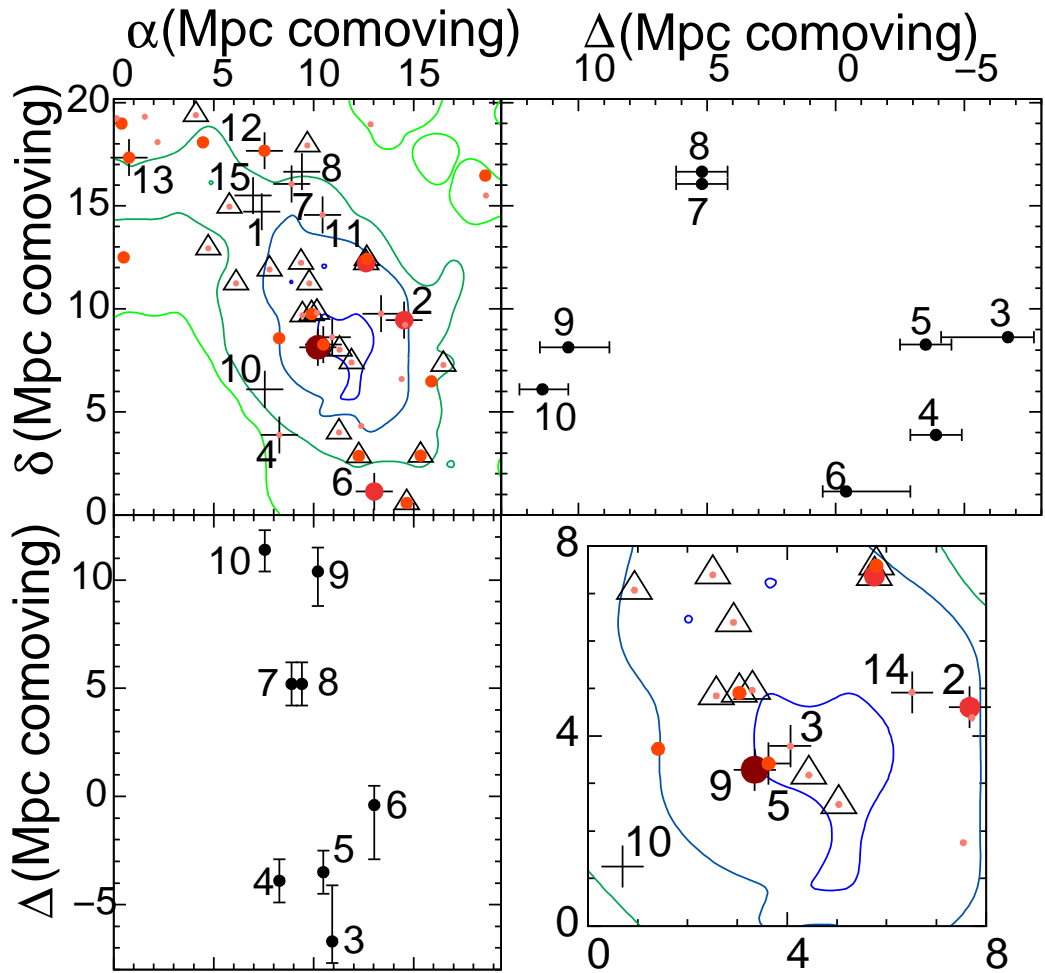}
\caption{Three-dimensional distribution of protocluster member galaxies.
    The upper left panel shows the spectroscopically observed region.
    The lines and points are the same as those in Figure \ref{cntr}, and the crosses and 
    triangles are spectroscopically observed objects, the Ly$\alpha$ emissions of which were detected 
    and not detected, respectively.
    The numbers near the points are the IDs (Column 1 of Table 3).
    The lower right panel is a closeup of the central region of the protocluster.
    The vertical axis in the lower left panel and the horizontal axis in the upper right panel, $\Delta$, 
    represent radial distance with an origin at $z=6$. 
    Note that in this figure, we regarded redshift difference as distance difference, although redshift
    difference might be caused by peculiar velocities inside the protocluster.\label{cluster}}
\end{figure*}

\subsection{Protocluster Properties}
We estimated the observed properties of the spectroscopically confirmed galaxies, properties such as the
Ly$\alpha$ fluxes, the UV continuum magnitudes, and the Ly$\alpha$ rest-frame equivalent widths (EWs).
Although faint continuum flux could not be detected in the observed spectra, the UV continuum magnitudes
were estimated from the $z'$-band magnitudes by subtracting the spectroscopically measured Ly$\alpha$ fluxes
and assuming flat UV continuum spectra ($f_\nu=\mathrm{constant}$).
The 15 $z\sim6$ galaxies are divided into two groups; the protocluster ``members,'' including ID=3-10,
and ``non-members,'' including ID=1, 2, and 11-15.
We compared the UV continuum luminosity, Ly$\alpha$ luminosity, rest-frame EW and FWHM of these two groups.
The average of these parameters between the members and the non-members agree with each other to within
$1\sigma$ scatter.
The composite spectrum of the members is shown in Figure \ref{comp}, and its weighted skewness is
$S_w=8.86\pm2.14$, whereas the weighted skewness of the composite spectrum of the non-members is
$S_w=8.70\pm2.04$.
Thus, the properties of the members are not found to be significantly different from those of non-members
despite the difference in the environments between the two samples.
However, we should note that the scatter in each of these properties is too large to conclude definitely
and that additional spectroscopic confirmations are required for more accurate comparisons.
It is also worth noting that we have not found apparent signatures of AGN activity for any of the member
galaxies.
At the center of the protocluster, there is a galaxy (ID=9) whose $z'$-band magnitude of $z'=25.47$ is
$>0.9\,\mathrm{mag}$ brighter than the average of all members.
This bright galaxy has an $\mathrm{SFR(UV)}=110.4\,\mathrm{M_\sun\,yr^{-1}}$, 
which seems too small for strong AGN activity.
We also did not detect any \ion{N}{5} $\lambda$1240{\AA} emission from the object.
This was the only accessible line that could have indicated AGN activity in our optical spectrum, 
although deeper spectroscopy is required to detect its AGN signature, if present.
In \S2.2, we mentioned one object that was too bright to meet the $4^\mathrm{th}$ criterion.
The brightness of this object indicates that it may be an AGN, but it is not located in the protocluster region.
The excess of AGNs or strong AGN activity were found in low redshift protoclusters, possibly resulting from
merging, which would be enhanced in such a high-density environment \citep[e.g.,][]{lehmer09,kuiper11}.
Unlike the low redshift protoclusters, strong AGN activity could not be found in our protocluster region,
implying that the member galaxies of this $z\sim6$ protocluster may be too young to have the requisite conditions
needed for significant AGN activity.

\subsection{The Mass Estimate of the Protocluster}
The eight clustering galaxies are spread over a $3\farcm5 \times 6\farcm5$ region, 
corresponding to $8.3 \times 15.5 \,\mathrm{Mpc^2}$.
The radial velocity dispersion, estimated simply from the dispersion of measured wavelength
of the Ly$\alpha$ emission line, is $\sigma_r = 647 \pm 124 \, \mathrm{km\,s^{-1}}$.
The error was estimated from the uncertainty of the wavelength of the Ly$\alpha$ emission line,
taking into account the wavelength resolution and the strong sky lines.
Furthermore, the resonant scattering of Ly$\alpha$ line could lend to additional error.
\citet{venemans07} found that the velocity dispersions of protoclusters at $z\sim2\mathrm{-}3$ and
$4\mathrm{-}5$ are $\sigma_r\sim500\mathrm{-}1000$ and 300$\,\mathrm{km\,s^{-1}}$, respectively.
The dark matter velocity dispersion is predicted to increase with cosmic time, based on a numerical
simulation of cluster evolution \citep[e.g.,][]{eke98}.
This is roughly consistent with observations at $z=2\mathrm{-}3$ and $4\mathrm{-}5$, 
and is predicted to reach to $\sigma_r \sim 200\,\mathrm{km\,s^{-1}}$ at $z\sim6$.
Our estimate of the velocity dispersion at $z\sim6$ is much larger than this.
We will discuss this discrepancy in the next subsection.
The radius of the protocluster is $r \sim 1 \, \mathrm{Mpc}$ in physical units.
Thus, the virial mass, $M_\mathrm{vir} \sim 3\sigma_r^2 r / \mathrm{G}$, where G is the gravitational 
constant and $r$ is the radius of the structure in physical units, would be 
$2.9^{+1.2}_{-1.0} \times10^{14}\,\mathrm{M_\sun}$, simply assuming that the protocluster
region was in virial equilibrium.
However, the protocluster is expected to be far from virial equilibrium.

Instead, we used another method to estimate the mass, specifically, $M=\bar{\rho}V(1+\delta_m)$ 
\citep{steidel98,venemans05}, where $\bar{\rho}$ is the current mean matter density of the universe, 
$V$ is the comoving volume of the protocluster and $\delta_m$
is the mass overdensity of the protocluster.
The mean density is $\bar{\rho}=4.1\times10^{10}\,\mathrm{M_\sun\,Mpc^{-3}}$ for the cosmological
parameters assumed in this paper.
The mass overdensity, $\delta_m$, is related to the galaxy overdensity $\delta_\mathrm{gal}$, 
through $1+b\delta_m = C(1+\delta_\mathrm{gal})$, where $b$ is the bias parameter and $C$ takes 
into account the redshift space distortions \citep{steidel98}.
Semi-analytical models \citep{baugh99,kauffmann99} and hydrodynamic simulations \citep{blanton00,yoshikawa01}
predict that the bias parameter will increase with increasing redshift.
The bias parameter is predicted to be $b=4.0\pm0.8$ for $z=5.7$ LAEs by \citet{orsi08} based on their
semi-analytical model in the $\Lambda$CDM universe. 
Thus, we assumed the bias parameter for LAEs at $z\sim6$ to be $b=4.0\pm0.8$.
It should be noted that the observed bias parameter of LAEs still has a large uncertainty.
\citet{ouchi03} derived the bias parameter to be $b=5.0$ for $z\sim4.8$ LAEs, and \citet{kovac07}
derived $b=3.7$ for $z\sim4.5$ LAEs, so these results should be interpreted with caution.
$C$ can be approximated by $C=1+f-f(1+\delta_m)^{1/3}$ \citep{steidel98}, where $f$ is the rate of growth of
perturbations at the redshift of the protocluster.
At high redshift, this value is close to 1 \citep{lahav91}.
If we adopt the volume of the protocluster as $V=2.3\times10^3\mathrm{\,Mpc^3}$, with six protocluster
galaxies being identified in the volume (this excludes ID=8, 10, which are the serendipitous identifications),
the protocluster number density is $n_\mathrm{cluster}=(2.6\pm1.1)\times10^{-3}\,\mathrm{Mpc^{-3}}$.
Using the mean surface number density (\S3.1), the field number density is 
$n_\mathrm{field}= (1.5\pm0.1)\times10^{-4}\,\mathrm{Mpc^{-3}}$.
These yield a galaxy overdensity of $\delta_\mathrm{gal} = n_\mathrm{cluster}/n_\mathrm{field}-1 = 16\pm7$,
corresponding to $\delta_m = 2.1\pm0.9$.
Therefore, the protocluster mass is $M=(2.9\pm0.8)\times10^{14}\,\mathrm{M_\sun}$.

Although it is interesting that the masses estimated from two different methods are consistent
with each other, each mass estimate has large uncertainties.
In the first method, we assumed virial equilibrium, which is almost certainly not true for such
a structure in the early phase of its evolution.
The second method retains a large uncertainty in estimating mass overdensity
from the galaxy number density.
Additionally, we regarded differences in redshift as difference in distance within the protocluster.
However, clustering structure may break the Hubble flow; in this case, the protocluster volume would
be overestimated.
Nevertheless, based on these mass estimations, the halo mass of the present-day descendant can be predicted
in a statistical sense using the extended Press-Schechter (EPS) formalism \citep{bond91,bower91}.
We find that the mass of $\sim 2.9 \times 10^{14}\,\mathrm{M_\sun}$ at $z=6$ will become the mass of
$(3.5\mathrm{-}5.8) \times 10^{15}\,\mathrm{M_\sun}$ (68\% range of distribution function) at $z=0$.
The predicted mass of the descendant is higher than the typical mass of local rich clusters
$\sim 9 \times 10^{14}\,\mathrm{M_\sun}$ and comparable to the most massive cluster
$\sim 4 \times 10^{15}\,\mathrm{M_\sun}$ \citep{wen10}; thus, the high concentrations at $z\sim6$ might
become one of the richest clusters in the local universe.
The theoretical work by \citet{overzier09a} predicted that the most massive halo masses are
$\sim10^{13}\;\mathrm{M_\sun}$ at $z\sim6$, and these evolved to rich clusters with halo masses of
$\sim10^{14}\mathrm{-}10^{15}\;\mathrm{M_\sun}$ in the local universe.
The derived mass, which is overestimated compared with the prediction, would imply two ideas: either the
protocluster is in a very early stage of cluster formation, which leads such a large velocity dispersion, or it
is a progenitor of an extremely massive cluster in the local universe.

\subsection{Protocluster Structure}
The estimated velocity dispersion of the protocluster is much larger than those at lower redshifts, and 
larger than the dispersion predicted by CDM models.
This discrepancy could be due to the distinguishing three-dimensional structure of the protocluster.
Figure \ref{cluster} shows the three-dimensional galaxy distribution of the protocluster, 
with possible substructures.
The spectroscopically confirmed galaxies tend to be distributed on the surface of a sphere
$\sim8\,\mathrm{Mpc}$ in radius, but a void containing few spectroscopically identified galaxies seems to
be present at the center of the sphere. 
The $z\sim6$ galaxy candidates in the void are generally faint in $z'$-band of $z'\ga26.5$,
in which the detection rate of Ly$\alpha$ emission is only 22\%, as described in \S4.
In contrast, the fraction of Ly$\alpha$ emitters among LBGs at the faint end of the LBG luminosity
function is generally higher than that of the bright end \citep{stark11}.
Although this result seems to contradict ours, Stark et al. made much deeper spectroscopic
observations than ours.
In our bright sample with $z'$-band magnitude ($25.0<z'<26.5$), the LAE fraction with
$\mathrm{EW}>25\mathrm{\AA}$ is $30\pm17\%$.
This is comparable, within the error, to that of \citet{stark11}, which is $20\pm8.1\%$.
If the same trend of the fraction in \citet{stark11} applied to our faint sample, about half of the
spectroscopically undetected galaxies with $z'>26.5$ are expected to have Ly$\alpha$ emissions with small EW
($<85\,\mathrm{\AA}$), which is difficult to detect in our observations.
Therefore, it is likely that this effect is the source of the void seen at the center of the protocluster.

Although it is not clear whether the $z\sim6$ galaxy candidates in the void region, including
spectroscopically undetected galaxies, are protocluster members or non-members, we now consider
two extreme scenarios.
First, we suppose that all $z\sim6$ galaxy candidates in the void region are real protocluster members, 
forming a large protocluster with a radius of $\sim8\,\mathrm{Mpc}$.
Second, we suppose that none of them are protocluster members, that the eight protocluster galaxies
are divided into small subgroups.
In the void region, we found 17 $z\sim6$ galaxy candidates without Ly$\alpha$ emission lines.
If their redshift distribution is similar to that of our spectroscopically confirmed samples 
(the histogram of Figure \ref{hist}), eight galaxies would be expected to be protocluster members.
The total number of protocluster members is 16 in this case, and this supports the first picture.
In contrast, if the redshift distribution of the void galaxies mimics a uniform distribution 
(the line of Figure \ref{hist}), only one galaxy is expected to be a protocluster member galaxy.
In this case, we would assume the second picture was accurate.
As discussed in \S4, the fraction of the real protocluster members in the overdense region is 
expected to be 0.5.
We cannot determine which hypothesis is likely based on the numerical simulation by \citet{overzier09a}, 
because the fraction of protocluster member galaxies has a large scatter.

In the first case, the total number of protocluster member galaxies would be 25 at most.
Galaxies located in the void are considered to be LBGs, either with weak Ly$\alpha$ emission, or with
no Ly$\alpha$ emission.
In contrast, LBGs with strong Ly$\alpha$ emission are predominantly distributed in the outer region 
of the protocluster.
In this case, our estimate of velocity dispersion of $647\,\mathrm{km\,s^{-1}}$, using only outer 
protocluster galaxies, is an overestimate; the actual velocity dispersion would likely be much smaller.
LBGs are generally older than LAEs \citep[cf.,][]{eyles07,stark09,ono10}.
If this is the case in this protocluster, inner galaxies in the protocluster may have evolved faster
than the outer galaxies, and may have already finished their early LAE phase.
Based on this picture, the radial dependence of galaxies' properties has already appeared even in the
$z\sim6$ protocluster.
While these results are suggestive, it would be interesting to obtain deeper spectroscopic observation of such
systems to definitively see whether such an environmental effect is already in place at such early epochs.
A similar feature was found in a protocluster at $z\sim5$ \citep{kashikawa07}, though it was not
distinguished from the possible negative feedback effect from the QSO at the center of the protocluster.

In the second case, spectroscopically confirmed galaxies distributed on the surface of a sphere seem to 
split into three subgroups: the $1^\mathrm{st}$, containing ID=3-6; the $2^\mathrm{nd}$, containing
ID=7 and 8; and the $3^\mathrm{rd}$, containing ID=9 and 10.
These subgroups might be in the process of forming a larger protocluster through merging.
In this case, the overall protocluster system is expected to be far from virial equilibrium; 
thus, the dynamical mass obtained from the velocity dispersion of eight LAEs was a significant overestimate.
The velocity dispersion of a subgroup including only four galaxies (ID=3-6 in Table 3)
is estimated to be $224\pm152\,\mathrm{km\,s^{-1}}$, which is in a good agreement with the CDM model prediction
at $z\sim6$.
The virial mass of the subgroup is $2.0^{+3.5}_{-1.8}\times10^{13}\,\mathrm{M_\sun}$, and the mass of the 
descendant is predicted to be $(3.0\mathrm{-}8.9)\times10^{14}\,\mathrm{M_\sun}$ by the EPS formalism.
\citet{ouchi05} found two protoclusters at $z\sim5.7$, and their velocity dispersions were
$\sim180$ and $150\,\mathrm{km\,s^{-1}}$.
We may be looking at a site of cluster formation prior to collapse, in which three small groups of galaxies 
begin to merge with each other.
Such a merger may trigger AGN activity at the center of the protocluster in the future.
\citet{kuiper11} found a merging protocluster at $z\sim2.2$, and suggested that a merger could enhance star 
formation and AGN activity.
Either scenario presents intriguing possibility.

We regarded the redshift difference as the difference in radial distance when discussing two scenarios above;
however, these differences might be caused by the peculiar velocity inside the protocluster.
More detailed investigations are necessary to reveal primitive cluster formation.
A series of redshift determinations for the galaxies in the void, derived from deep spectroscopy, is the 
most straightforward approach for distinguishing between these two ideas.
If the first case is true, galaxies in the void region would have older ages, higher stellar masses, 
higher metallicities, and larger amounts of dust extinction.
We could confirm those points with observations taken at longer wavelengths to trace the rest-optical 
frame of $z\sim6$ galaxies, followed by accurate SED fitting.
We plan to address this issue in detail in the future.
If the second case is true, it would be interesting to investigate the possibility of additional subgroups
further outside the protocluster.
Spectroscopic observations designed to locate outer galaxies should reveal additional subgroups or 
filamentary structures, closely related to the initial large-scale structure of the universe.

\section{CONCLUSIONS}
In this paper, we have presented the discovery of a protocluster at $z\sim6$ in the SDF based on eight
spectroscopically confirmed member galaxies.
This discovery was achieved in a ``random field'' using wide-field imaging; that is, QSOs or RGs were
not used as protocluster probes.
The $i'$-dropout surface number density of the protocluster region is overdense at the $6\sigma$ level
compared with the mean density of the whole SDF.
The size of the overdense region is $\sim6\arcmin\times6\arcmin \,(14\,\mathrm{Mpc}\times14\,\mathrm{Mpc})$.
The protocluster galaxies are distributed in a narrow redshift range of $\Delta z < 0.05$ around $z=6.01$.
The concentration in redshift space is seven times higher than that expected from a uniform distribution.
This region is certainly a protocluster at $z=6.01$, the highest redshift such a structure has ever
been observed.
The differences in galaxy properties, such as Ly$\alpha$ luminosity and UV continuum magnitude, between
protocluster members and non-members (i.e., field galaxies) at $z\sim6$ cannot be detected, although
the small sample size still prevents us from forming a definite conclusion.

The estimated velocity dispersion of the protocluster is $\sigma_r = 647 \pm 124 \, \mathrm{km\,s^{-1}}$.
Assuming virial equilibrium, the protocluster mass is estimated to be
$2.9^{+1.2}_{-1.0} \times10^{14}\,\mathrm{M_\sun}$.
Based on the EPS formalism, we expect that, if virialized, this protocluster will become one of the richest
clusters ($\sim4.7\times10^{15}\,\mathrm{M_\sun}$) in the local universe.
From the three-dimensional distribution of the eight protocluster galaxies, we proposed
two different scenarios for the protocluster structure: either the protocluster is already mature, with
old galaxies at the center, or it is still immature and composed of three subgroups merging to become
a larger cluster.
Further deep spectroscopic observations or multi-wavelength imaging to trace the rest-frame optical
wavelengths would provide a clearer picture of the protocluster structure, as well as providing constraints
for the galaxy/stellar populations of protocluster members.
A statistical sample of $z\sim6$ protoclusters based on wider surveys would allow a direct comparison of
the observed protocluster abundance in the early universe with the predictions of the $\Lambda$CDM model.

\acknowledgments
This paper is based on data collected with the Subaru Telescope, which is operated by the National
Astronomical Observatory of Japan.
We are grateful to the Subaru Observatory staff for their help with the observations.
We thank the anonymous referee for valuable comments and suggestions which improved the manuscript.
This research was supported by Japan Society for the Promotion of Science through 
Grant-in-Aid for Scientific Research 23340050.

{\it Facilities:} \facility{Subaru (SuprimeCam, FOCAS)}, \facility{Keck:II (DEIMOS)}, \facility{UKIRT (WFCAM)}

\begin{figure*}
\epsscale{1.0}
\plotone{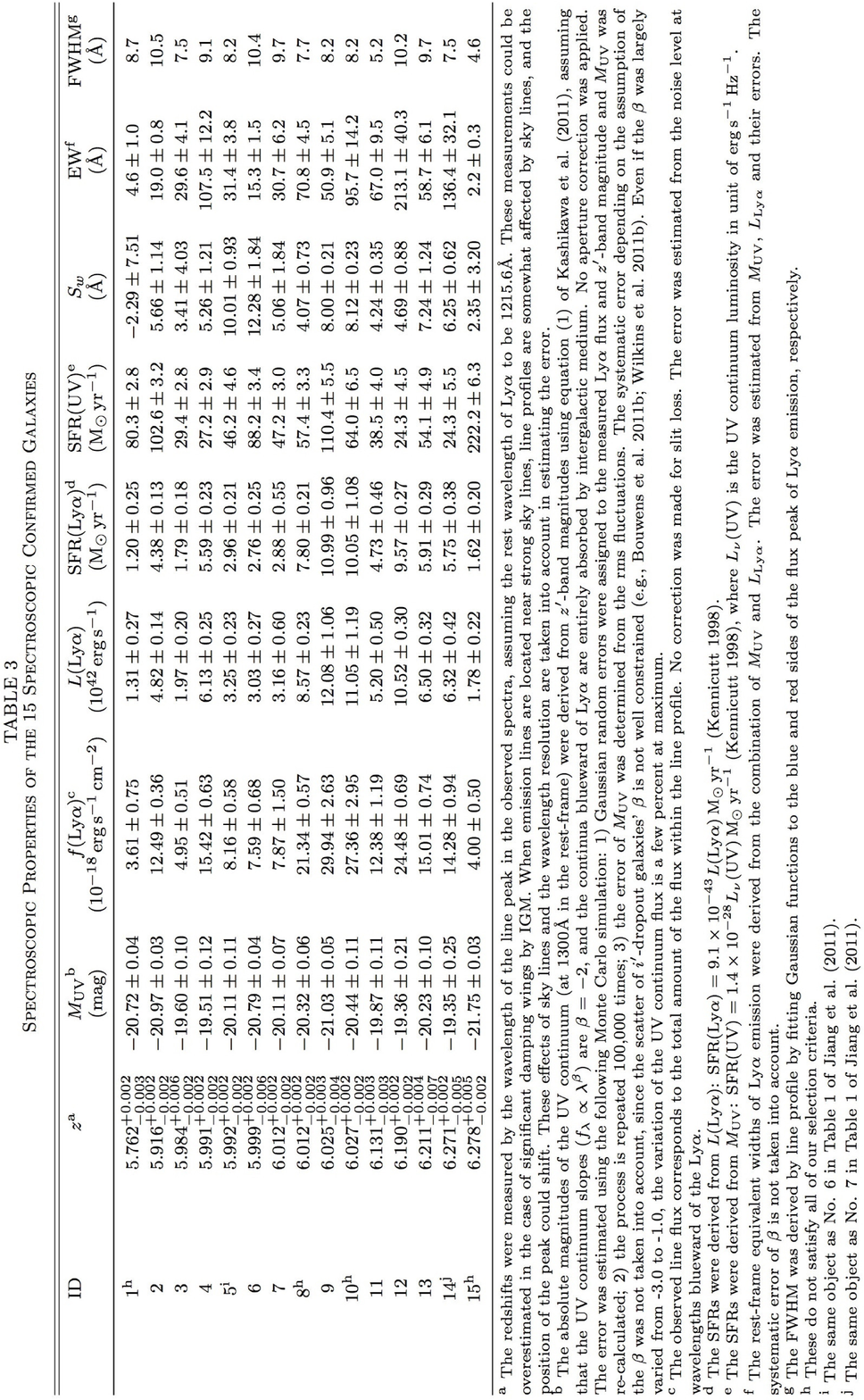}
\end{figure*}

\end{document}